\documentclass[fleqn,twoside]{article}
\usepackage{espcrc2}

\usepackage{graphicx}
\usepackage{epsfig}
\usepackage[figuresright]{rotating}

\newcommand{\AmS}{{\protect\the\textfont2
  A\kern-.1667em\lower.5ex\hbox{M}\kern-.125emS}}

\newcommand{\be}{\begin{equation}}
\newcommand{\ee}{\end{equation}}
\newcommand{\bi}{\begin{itemize}}
\newcommand{\ei}{\end{itemize}}
\newcommand{\ben}{\begin{enumerate}}
\newcommand{\een}{\end{enumerate}}
\newcommand{\bc}{\begin{center}}
\newcommand{\ec}{\end{center}}
\newcommand{\bea}{\begin{eqnarray}}
\newcommand{\eea}{\end{eqnarray}}
\newcommand{\nn}{\nonumber}

\def      \raw           {\rightarrow}

\def      \nue{\ensuremath{\nu_{e}}\ }

\def      \numu{\ensuremath{\nu_{\mu}}\ }

\def      \nutau{\ensuremath{\nu_{\tau}}}

\def      \simge{\mathrel{%
                \rlap{\raise 0.511ex \hbox{$>$}}{\lower 0.511ex \hbox{$\sim$}}}}
\def      \simle{\mathrel{
                \rlap{\raise 0.511ex \hbox{$<$}}{\lower 0.511ex \hbox{$\sim$}}}}
\def      \BB{$\beta$-Beam\ }

\def      \NF{$\nu$-Factory\ }

\title{Physics reach of $\beta$-beams and $\nu$-factories: the problem of degeneracies}

\author{S. Rigolin\address[IFT]{Theoretical Physics Department and I.F.T, 
        Universidad Autonoma de Madrid, Cantoblanco, Spain}%
        \thanks{The author acknowledges the financial support of MCYT through project 
        FPA2003-04597 and of the European Union through the networking activity BENE.}}
       
\begin{document}

\begin{abstract}
We discuss the physics reach of $\beta$-Beams and $\nu$-Factories from a theoretical 
perspective, having as a guideline the problem of degeneracies. The presence of degenerate 
solutions in the measure of the neutrino oscillation parameters $\theta_{13}$ and $\delta$ 
is, in fact, the main problem that have to be addressed in planning future neutrino oscillation 
experiments. If degeneracies are not (at least partially) solved, it will be almost impossible 
to perform, at any future facility, precise measurements of $\theta_{13}$ and/or $\delta$. 
After a pedagogical introduction on why degenerate solutions arise and how we can get rid of 
them, we analyze the physics reach of current $\beta$-beam and $\nu$-factory configurations. 
The physics reach of the "standard" \BB is severely affected by degeneracies while a better 
result can be obtained by higher-$\gamma$ setups. At the \NF the combination of Golden and 
Silver channels can solve the eightfold degeneracy down to $\sin^2\theta_{13} \le 10^{-3}$. 
\vspace{1pc}
\end{abstract}

\maketitle

\section{Introduction}
\label{introduction}

The atmospheric and solar sector of the PMNS leptonic mixing matrix have been measured 
with quite good resolution by SK, SNO and KamLand. These experiments measure two angles, 
$\theta_{12}$ and $\theta_{23}$, and two mass differences, $\Delta m^2_{12}$ and 
$\Delta m^2_{23}$. The present bound on $\theta_{13}$, $\sin^2 \theta_{13} \leq 0.04$, 
is extracted from the negative results of CHOOZ and from three-family analysis of 
atmospheric and solar data. The PMNS phase $\delta$ is totally unbounded as no experiment 
is sensitive, up to now, to the leptonic CP violation. The main goal of next neutrino 
experiments will be to measure these two, still unknown, parameters. In this talk we 
analyze the physics reach of \BB and \NF following a somehow theoretical perspective: the 
problem of degeneracies in ($\theta_{13},\delta$) measure.
The best channel for measuring $(\theta_{13},\delta)$ is the $\nue \!\!\raw \numu$ 
appearance channel \cite{Cervera:2000kp} (and/or its T and CP conjugate ones). Unfortunately,
this measure is severely affected by the presence of an eightfold degeneracy 
\cite{Burguet-Castell:2001ez,degeneracies}. 

In the next section we introduce in a pedagogical way the concept of degeneracy and we 
describe the effect of the eightfold degeneracy in the $(\theta_{13},\delta)$ measure. We 
show that two possible approaches can be used to get rid of them: combination of 
different experiments and/or combination of different oscillation channels. In section 3 
and 4 we describe shortly the physics reach of \BB and \NF respectively\footnote{The 
eightfold degeneracy in the ($\theta_{13},\delta$) measure has been comprehensively 
studied in literature. All the technical details and a complete set of bibliographic 
references can be found in \cite{Donini:2003vz,Donini:all}}. 

\section{The eightfold degeneracy}
\label{theo}

\begin{figure}[t!]
\hspace{-0.5cm}
\epsfig{file=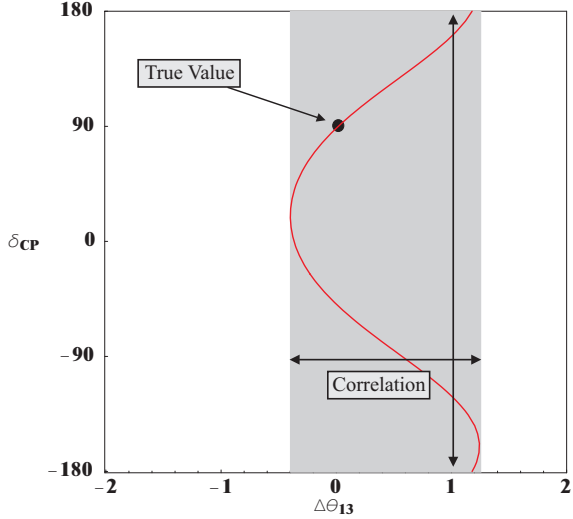,width=7.5cm,angle=0}\vspace{-0.5cm}
\caption{Correlation of $\theta_{13}$ and $\delta$ if only neutrinos (or antineutrinos) 
are measured: infinite degeneracy.}
\label{fig:deg0}
\end{figure}

\begin{figure}[t!]
\hspace{-0.5cm}
\epsfig{file=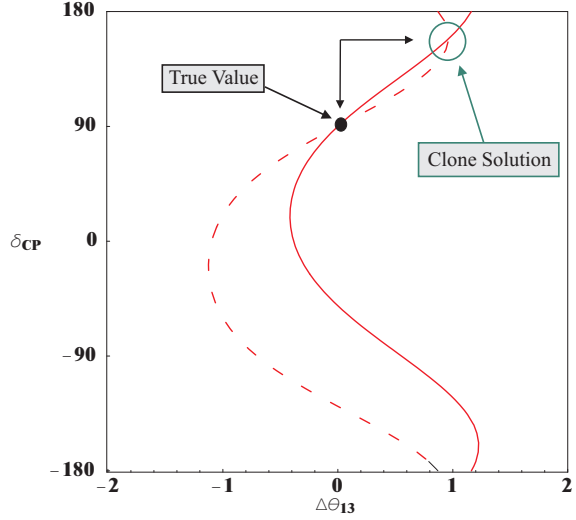,width=7.5cm,angle=0}\vspace{-0.5cm}
\caption{Correlation of $\theta_{13}$ and $\delta$ if neutrinos (full line) and antineutrinos 
(dashed line) are measured: twofold degeneracy.}
\label{fig:deg1}
\end{figure}

It was originally pointed out in Ref.~\cite{Burguet-Castell:2001ez} that the appearance 
probability $P_{\alpha \beta}$ for neutrinos with a fixed Baseline/Energy (L/E) ratio and 
input parameters ($\bar\theta_{13},\bar\delta$) has no unique solution. Indeed, the equation 
\be
\label{eq:equi0}
P_{\alpha\beta} (\bar\theta_{13},\bar\delta) = P_{\alpha\beta}
(\theta_{13},\delta)
\ee
has a continuous number of solutions. The~locus of the ($\theta_{13},\delta$) plane
satisfying this equation is called {\em equiprobability curve} (see Fig.~\ref{fig:deg0}). 
%
%
Consider now the equiprobability curves for neutrinos ($+$) and antineutrinos ($-$) with 
the same L/E (and the same input parameters). The following system of equations
\be
\label{eq:equi1}
P^\pm_{\alpha \beta} (\bar \theta_{13},\bar \delta) = P^\pm_{\alpha \beta}
(\theta_{13}, \delta)
\ee
has two intersections (see Fig.~\ref{fig:deg1}): the input pair ($\bar \theta_{13},\bar \delta$) 
and a second, L/E dependent, point. This second intersection introduces an ambiguity in the 
measurement of the physical values of $\theta_{13}$ and $\delta$: the so-called {\it intrinsic 
clone} solution \cite{Burguet-Castell:2001ez}. 
Knowing the two probabilities of eq.~(\ref{eq:equi1}) is consequently not enough for solving 
the intrinsic degeneracy. One needs to add more informations. 
%
%

\begin{figure}[t!]
\hspace{-0.5cm}
\epsfig{file=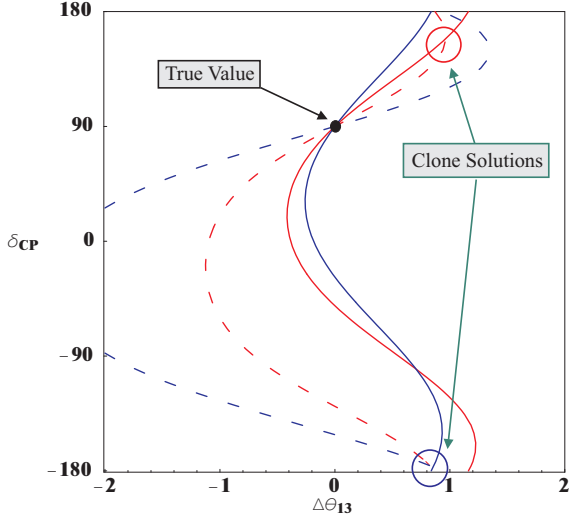,width=7.5cm,angle=0} \vspace{-0.5cm}
\caption{Solving the intrinsic degeneracy using the same oscillation channel but two 
different values of L/E.} 
\label{fig:deg2}
\end{figure}

\begin{figure}[t!]
\hspace{-0.5cm}
\epsfig{file=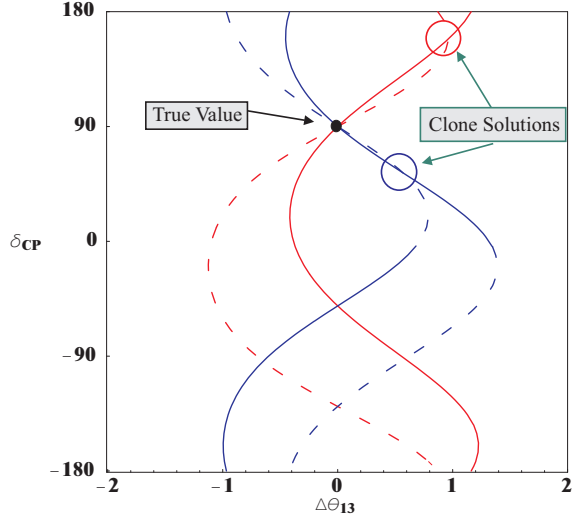,width=7.5cm,angle=0} \vspace{-0.5cm}
\caption{Solving the intrinsic degeneracy using the same L/E but two different oscillation 
channels (i.e. golden and silver).} 
\label{fig:deg3}
\end{figure}

Two ways are viable: i) using independent experiments (i.e different L/E) and/or ii) 
using independent oscillation channels. In case i) one can think to observe the same 
neutrino oscillation channel (i.e. the golden $\nu_e \raw \nu_\mu$ oscillation) using 
neutrino (antineutrino) beams with different L/E. In Fig. \ref{fig:deg2} one can see 
that experiments with different L/E present clone solutions in different regions of 
the ($\theta_{13},\delta$) parameter space. If the clones are well separated one can 
solve the degeneracy. 
%
Another possibility, case ii), is to fix L/E and to use contemporaneously two different 
oscillation channels (like for example $\nu_e\raw \nu_\mu$ and $\nu_e \raw \nu_\tau$). 
In Fig. \ref{fig:deg3} one can see how the intrinsic clones for the two channels appear, 
generally, in different locations and so the intrinsic degeneracy can be solved. 

From this example we learn that the best way for solving the degeneracies is to add all 
the possible available informations: different baselines, different energy bins (i.e. 
different L/E) and different channels. Therefore, in planning future experiments one has 
to understand which combinations of experiments can give the largest set of (really) 
independent informations. The existence of unresolved degeneracies could, in fact, 
manifests itself in a complete lost of predictability on the aforementioned parameters 
as we will see in the next sections.

Unfortunately, the appearance of the intrinsic degeneracy is only a part of the 
``clone problem''. As it was pointed out in \cite{degeneracies}, two other sources of 
ambiguities arise due to the present (and near future) ignorance of the sign of the 
atmospheric mass difference, $s_{atm}=~\mbox{sign}[\Delta m^2_{23}]$, and of the 
$\theta_{23}$ octant, $s_{oct}=~\mbox{sign} [\tan(2\theta_{23})]$. These two discrete
variables assume the values $\pm 1$, depending on the physical assignments of the 
$\Delta m^2_{23}$ sign ($s_{atm}=1$ for $m_3^2>m_2^2$ and $s_{atm}=-1$ for $m_3^2<m_2^2$) 
and of the $\theta_{23}$-octant ($s_{oct}=1$ for $\theta_{23}<\pi/4$ and $s_{oct}=-1$ 
for $\theta_{23}>\pi/4$). Consequently, future experiments will have to measure four 
unknowns: two continuous variables ($\theta_{13}$,$\delta$) plus two discrete 
variables ($s_{atm}$,$s_{oct}$).

From these considerations it follows that eq.~(\ref{eq:equi1}) should be more correctly 
replaced by the following four systems of equations (each for any possible choice of 
the $s_{atm}$ and $\bar s_{oct}$ signs)\footnote{For simplicity of notation we express 
eqs.~\ref{eq:equi0}--\ref{eq:ene0int} in terms of probabilities. However as noticed in 
\cite{Donini:2003vz} one should use instead the number of (leptonic) measured events.}: 
\bea
\mathindent=0pt
\label{eq:ene0int}
P^\pm_{\alpha \beta} (\bar \theta_{13},\bar \delta; \bar s_{atm},\bar s_{oct}) = & & \nn \\
              & & \hspace{-3.4cm} 
P^\pm_{\alpha \beta} (\theta_{13},\delta; s_{atm}=\pm \bar s_{atm}; s_{oct}=\pm \bar s_{oct}) . 
\eea
\noindent
Solving the four systems of eq.~\ref{eq:ene0int} will result in obtaining the true 
solution plus additional {\it clones} to form an eightfold degeneracy. These eight 
solutions are respectively: the true solution and its {\em intrinsic clone} (when the 
right $s_{atm}$ and $\bar s_{oct}$ signs are used in eq.~\ref{eq:ene0int}), the 
{\em $\Delta m^2_{23}$-sign clones} (when $s_{atm}=-\bar s_{atm}$ is used), the 
{\em $\theta_{23}$-octant clones} (when $s_{oct}=-\bar s_{oct}$ is used) and finally the 
{\em mixed clones} (when simultaneously $s_{atm}=-\bar s_{atm}$ and $s_{oct}=-\bar s_{oct}$ are used).

%
\section{Physics reach of the \BB}
\begin{figure}[t!]
\hspace{-0.5cm}
\epsfig{file=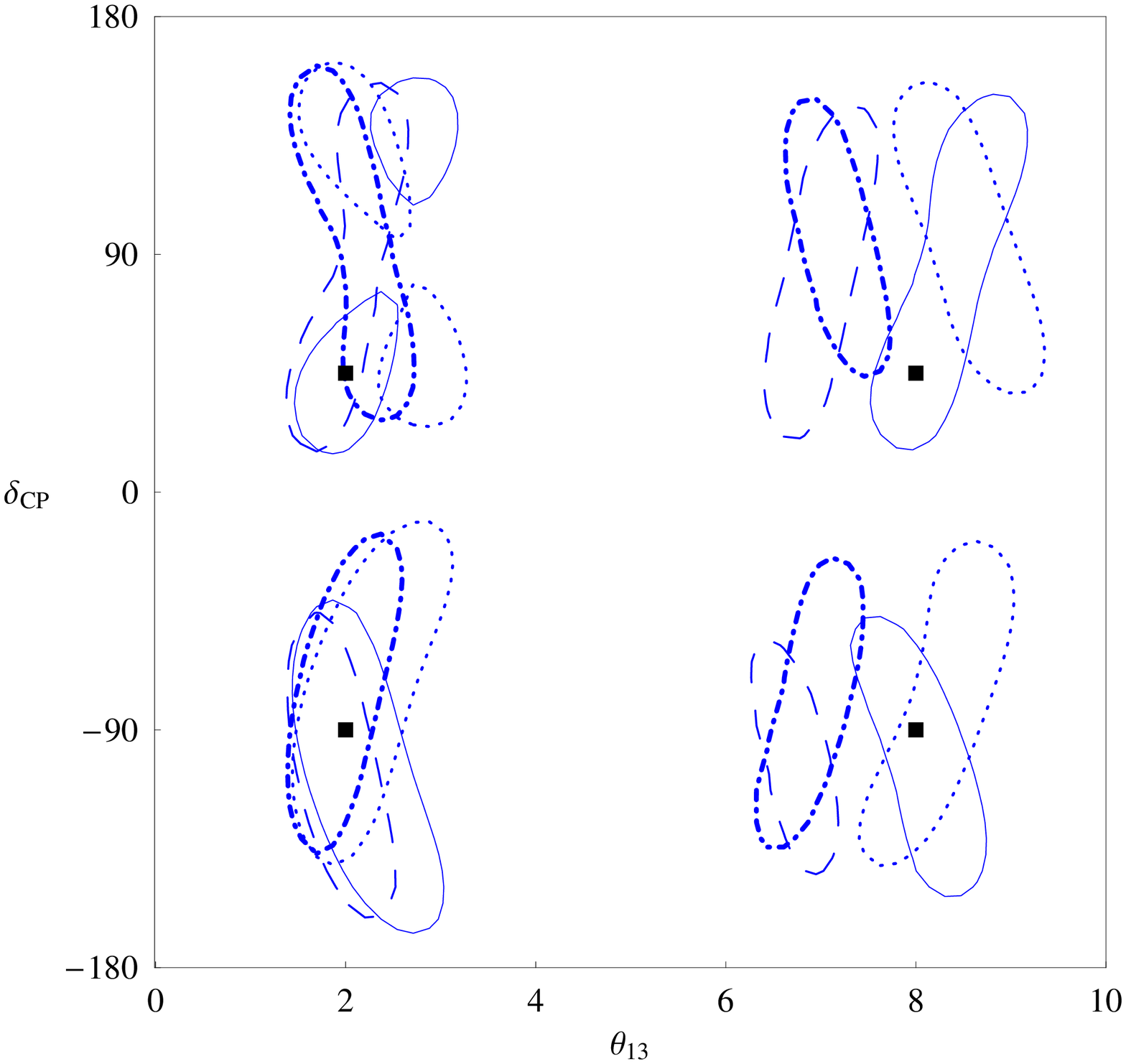,width=7.5cm,angle=0} \vspace{-0.5cm}
\caption{The eightfold degeneracy at the ``standard" \BB for two values of $\theta_{13}$ ($2^\circ, 
8^\circ$) and $\delta$ ($45^\circ, -90^\circ$) (from \cite{Donini:2004iv}).} 
\label{fig:BB1}
\end{figure}

In Fig.~\ref{fig:BB1} one can see the dramatic impact that degeneracies can have 
in the precision measurement of $(\theta_{13},\delta)$ at the ``standard" \BB 
configuration\footnote{For a comprehensive description of old and new \BB setups 
look at \cite{Donini:all,Donini:2004iv,allBB,Burguet-Castell:2005pa}} (i.e. L=130 km 
and $\gamma$=(60,100) for He and Ne ions respectively): 
(1) the error in the $\theta_{13}$ measurement is increased by a factor four (two) 
    for large (small) values of $\theta_{13}$, however, the presence of degeneracies has 
    a small impact on the ultimate $\theta_{13}$ sensitivity; 
(2) the error in the $\delta$ measurement grows in a significant way in presence of the 
    clones, almost spanning half of the parameter space for small values of $\theta_{13}$. 
These facts are well understood: being the ``standard" \BB a (short distance) counting 
experiment there are not enough independent informations to cancel any of the degeneracies. 

%
\begin{figure}[t!]
\hspace{-0.25cm}
\epsfig{file=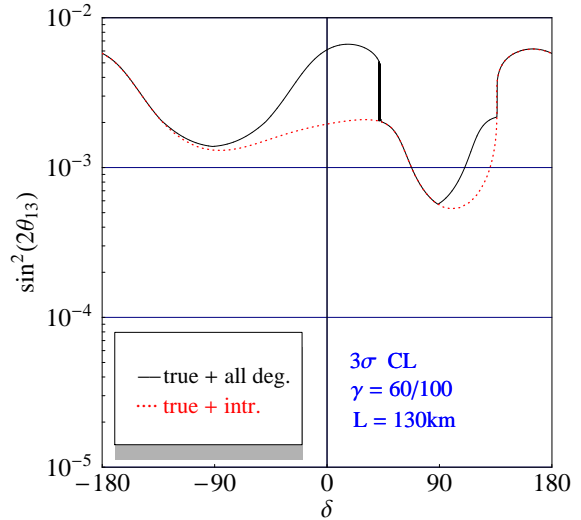,width=7.5cm,angle=0} \vspace{-0.5cm}
\caption{$\theta_{13}$ exclusion plot for the ``standard" \BB setup 
(from the authors of \cite{Donini:2004iv}).} 
\label{fig:BB2}
\end{figure}

To obviate this problem new \BB setups have been recently proposed by 
\cite{Burguet-Castell:2005pa,Huber:2005jk}. In these setups the use of energy bins 
helps in reducing the number of degeneracies and improves the sensitivity to $\theta_{13}$ 
and $\delta$. However, to have a significant improvement over the ``standard" setup, 
neutrino energies of ${\cal O}$(1 GeV) (i.e. $\gamma \approx$ 150-300) and baselines 
of 300-700 km are needed.  

In Fig.~\ref{fig:BB2} the 3$\sigma$ sensitivity reach to $\theta_{13}$ is shown. 
Notice that the inclusion of all degeneracies worsen the sensitivity to $\theta_{13}$ 
for $\delta$=0 of almost and order of magnitude, compared to the only-intrinsic case 
usually presented in literature. 

\begin{figure}[t!]
\hspace{-0.6cm}
\epsfig{file=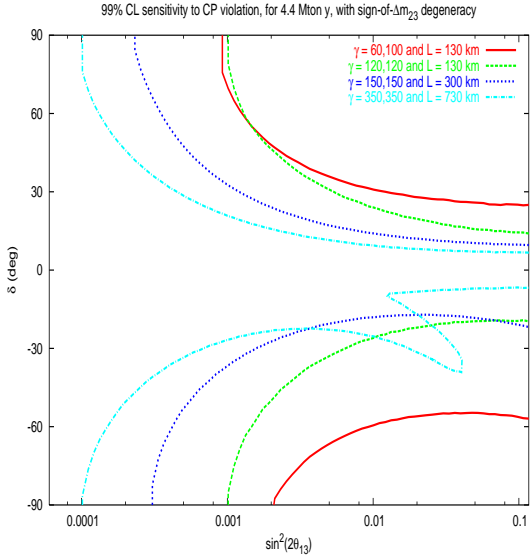,height=7.25cm, width=7.5cm,angle=-90} \vspace{-0.5cm}
\caption{CP violation exclusion plot at 99\% CL for different \BB setups (from 
ref. \cite{Burguet-Castell:2005pa}).} 
\label{fig:BB3}
\end{figure}

In Fig.~\ref{fig:BB3} the CP violation exclusion plot at 99\% CL for different \BB setups 
it is shown \cite{Burguet-Castell:2005pa}. One can easily notice in this plot that the 
sensitivity to CP violation is increased at higher-$\gamma$ (and baseline) configurations. 
This work, however, assumes the same neutrino flux at any of the $\gamma$ considered. While 
this hypothesis seems reasonable (within a factor 2), a careful analysis of ions 
fluxes at different $\gamma$s should be performed.

Besides the appearance channel, the $\nue \!\! \raw \nue \!\!$ disappearance channel is 
available at the $\beta$-Beam. Unfortunately, as it was shown in \cite{Donini:2004iv} this 
channel does not provide any additional informations once realistic systematic errors are 
taken into account. 


%
%
\section{Physics reach of the \NF}

%

\begin{figure}[t!]
\begin{center}
\hspace{-0.25cm}
\epsfig{file=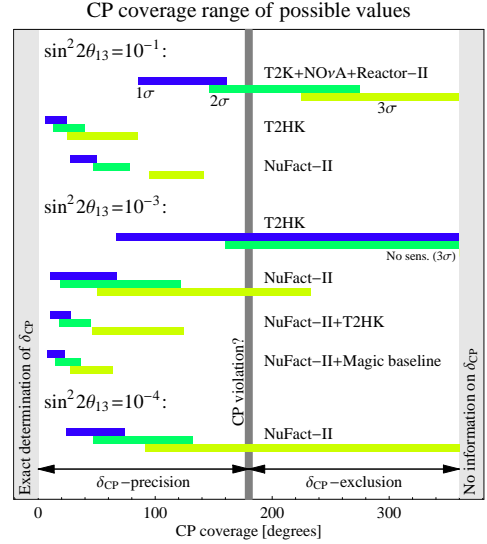,height=7.2cm,angle=0} 
\vspace{-0.6cm}
\caption{CP coverage summary (from the last paper in ref.~\cite{Burguet-Castell:2002qx}).} 
\label{fig:CPcoverage}
\end{center}
\vspace{-0.5cm}
\end{figure}

No much work has been devoted in the last two years in improving the analysis for the 
$\nu$-Factory. 
Our understanding of the \NF physics reach is practically still the one presented in 
ref.~\cite{Burguet-Castell:2002qx} (and graphically summarized in Fig.~\ref{fig:CPcoverage})
with the addition of the $\nue \!\! \raw \nutau \!\!$ ``silver" appearance channel 
ref.~\cite{Donini:2002rm}. The combination of the ``golden" channel (with a 3000 km baseline 
and, eventually, a second detector placed at 7000 km) and the ``silver" channel could solve 
practically all the degeneracies down to $\theta_{13} \approx 1^\circ-2^\circ$. Notice 
however, that all the golden channel studies strongly dependend on the 
old-fashion\footnote{Originally the detector was designed for reaching the highest sensitivity 
to $\theta_{13}$ because at the time the detector was designed the SMA solution was not 
excluded yet.} detector configuration presented in \cite{Cervera:2000vy}. The appearance 
of the "degeneracies problem" force us to consider a complete re-analysis of the iron detector 
characteristics. The inclusion of the first energy bins (let's say between 0-10 GeV) becomes
mandatory for solving clones solutions and so improving \NF physics reach. In fact the 
oscillation maximum for a NF with a baseline of 3000 km is around 5 GeV. In the old detector 
analysis tight background cuts reduce almost to 0 the efficiency for this bin, in such a way 
that only the above-the-peak energy informations is left.

In our opinion, the full understanding of the \NF physics reach should have to take advantage 
of its large availability of oscillation channels. Besides the accurate studies of the 
``golden" and ``silver" appearance channels, a deeper analysis of the $\nue,\numu$ disappearance 
channels and of the $\numu \!\!\raw \nutau$ appearance one should be considered. Of course this 
effort could require a (considerable) increase of detectors cost. But it should be considered 
if the major effort of building the \NF is agreed to be necessary for the full understanding 
of neutrino oscillation parameters. 

%
%

\clearpage
\end{document}